\documentclass[10pt]{article}
\usepackage{amsfonts,latexsym}

\usepackage{times}
\usepackage{graphics}

\newcommand{\ie}{{\it i.e.}}
\newcommand{\eg}{{\it e.g.}}
\newcommand{\Z}{\mathbb{Z}}
\newcommand{\R}{\mathbb{R}}
\newcommand{\be}{\begin{equation}}
\newcommand{\ee}{\end{equation}}
\newcommand{\Ref}[1]{(\ref{#1})}
\newcommand{\rep}[1]{\sigma^{(#1)}}
\newcommand{\ba}{\begin{array}}
\newcommand{\ea}{\end{array}}  
\newcommand{\bea}{\begin{eqnarray}}
\newcommand{\eea}{\end{eqnarray}}  
\newcommand{\dlangle}{\langle\hspace{-.6ex}\langle}
\newcommand{\drangle}{\rangle\hspace{-.6ex}\rangle}
\newcommand{\sign}{\mathrm{sign}}
\newcommand{\vol}{\textbf}

\newcommand{\plaq}{\Box} 

\begin{document}
\begin{center}
{\Large 
Evidence for Frustration Universality Classes \\ in 3D Spin Glass
Models ~\footnote{This work is supported in part by ``Enrico
  Fermi Research Center'', via Panisperna 89, Rome.}\\
}

\bigskip

{\large Leonardo Gnesi, Roberto Petronzio, Francesco Rosati}\\

\bigskip
{\it
Dipartimento di Fisica, Universit\`a degli Studi di Roma ``Tor
Vergata'',\\ and I.N.F.N.\ Sezione Roma II,
via della Ricerca Scientifica, 1, I-00133 Roma, Italia
}
\bigskip
\bigskip
\bigskip

\begin{minipage}{0.8\textwidth}
{
\small
\begin{center}  {\bf Abstract} \end{center}
\noindent
A recently introduced Renormalization Group approach to frustrated
spin models is applied in three dimensions through
Monte Carlo computations. A class of spin glass models is analysed,
with correlated disorder variables given by a $\Z_2$ gauge
field. Evidence is provided for the influence of deconfinement phase
transition of gauge fields on the behaviour of the associated spin
models, namely in the transition from glassy to ferromagnetic
behaviour. Universality classes are determined on fundamental ground
by characterizing the fixed point. Moreover, the RG analysis provides
good estimates of the critical temperature and the thermal index
$\nu$, with moderate computer time.

Renormalization group methods.
}
\end{minipage}
\end{center}

\bigskip

\bigskip
\section{Introduction}

The renormalization group (RG) is an invaluable tool in the understanding
of critical phenomena~\cite{Jona:rassegna,Wilson71}. It provides accurate
estimates of critical indexes as well as a theoretical foundation of the
universality principle and of the scaling theory. Many
authors~\cite{Wang:1988,droplet,Palassini:1999} have profitably used these ideas in
spin glasses (SG's), even if a general RG framework for SG's was incomplete. 

In a previous paper~\cite{NOI} a RG approach suitable
for SG's has been proposed and fruitfully exploited through Monte Carlo (MC)
computations in dimensions $d=4$. A sketch of expected universality classes and
the role of frustration were discussed. 
In the present work, these ideas has been fully developed
in the more relevant and
harder case of $d=3$. 
The coarse graining transformation is carried out on the \emph{overlap} 
field. Because of disorder average, overlap probability measure 
does not have the usual form $\exp(-H)$ , hence it is not \emph{a priori}
clear which kind of additional interactions are generated by  the RG transformation.
Our proposal is to keep Hamiltonian fixed while adding gauge
invariant terms to the disorder distribution, so that spins get
effective correlations
via correlations of gauge fields. Asymptotic
decrease regimes of Wilson loops in pure gauge model~\cite{Toulouse:1977}, namely the
\emph{weak} and \emph{strong} decrease, can induce very different behaviour
on related spin systems. In the former case, frustration effects are
very small locally and increase weakly at long range: spins
prefer to order ferromagnetically. In the case of strong decrease,
instead, average Wilson loop rapidly fall to zero and a SG behaviour is
expected. The existence of two fixed points corresponding to the above mentioned
universality classes is supported by evidence from our
analysis. Moreover, our estimates of critical temperature $T_\mathrm{c}$ and
thermal index $\nu$ are in good agreement with previous results~\cite{Ballesteros:2000,Pellis}.
Notice that a less frustrated system gets ordered more easily,
giving rise to higher critical temperatures. This property can be
exploited to perform Monte Carlo computations of SG models at very low
$T/T_\mathrm{c}$ ratios~\cite{GPR}. 

Numerical methods based on Monte Carlo simulations of SG's are
particularly hard in $d=3$ basically because of the 
low temperatures involved. This requires
very long Monte Carlo runs to reach ergodic regime. 
Parallel Tempering (PT) algorithm, which we implement, 
is adequate to face this kind of problems but
unfortunately it hides the meaning of MC autocorrelation time, making
standard error estimates not efficient. Aiming to control both
ergodicity and MC-errors, we employ a method based on the
analysis of MC-PT dynamics, which turns out to be quantitative and
more stringent than usual PT efficiency checks.

\section{The spin glass model and the overlap field}\label{sec:SG}

Let $\sigma_x$ be
Ising spins located at the sites of a $d$-dimensional cubic lattice
$\Lambda$, with $L$ points on each side ($x\in \Lambda\subset\Z^d$
with toroidal topology, \ie, periodic boundary conditions are assumed).
The typical spin glass model of Edwards
and Anderson~\cite{EA75} is defined by two prescriptions: 1. The spins are 
distributed according to the Boltzmann weight given by the following
Hamiltonian
\be
H(J,\,\sigma) = - \sum_{\langle x,\,y\rangle} J_{xy}\sigma_x\sigma_y
;\label{hamiltonian}
\ee
2. The quenched disordered interactions
$J_{xy}$ are random variables, with a given distribution. 
The sum in \Ref{hamiltonian} is over pairs of neighboring sites in
$\Lambda$. We will denote by $E$ the expectation on $J$ variables, and
define 
\be 
E(J_{xy}) = 0,\;\;E(J_{xy}^2) = 1
.\label{EdiJ}
\ee
The Boltzmann--Gibbs measure on the spin variables will be denoted by
angular brackets~$\langle\cdot\rangle$. Averages over the disorder
are taken only after Boltzmann averages are calculated,
and the thermodynamic limit for the appropriate quantities 
is eventually taken afterwords. 
A generic observable can be expressed in terms of correlations {\it of}
correlation functions,
\be 
E\left[ 
\langle f_1(\sigma) \rangle\ldots\, \langle f_s(\sigma) \rangle
\right]
\label{observ}
,\ee
where $f_i,\,i=1,\ldots,s$ are functions of the spin field.
Consider the following gauge symmetry of the Hamiltonian,
\be
\left\{ \ba{l}
 J_{xy}\rightarrow J'_{xy} = \varepsilon_x J_{xy} \varepsilon_y \\
 \sigma_x\rightarrow \sigma'_x = \varepsilon_x\sigma_x \\
\ea \right.
\ee
where $\varepsilon_x = \pm 1$ are the gauge group parameters.
If the distribution of the $J$'s is gauge-invariant, the model,
characterized by observables \Ref{observ}, is also gauge-symmetric.
This happens, for instance, in the traditional case of even,
independent, identically distributed $J$ variables (E-A model). 

A central concept is the overlap field. Consider $s$ replicas
(copies) of the spin variables $\rep{a},\, a=1,\ldots,s$. 
The Hamiltonian of the replicated system is given by 
${\mathcal H} = \sum_a H(J,\,\rep{a})$: replicas are
independent from each other, but feel the {\it same} disorder
configuration $J$. For any pair of replicas $(a,b),\; 1\leq a<b\leq
s$, define the {\it overlap field}, $q^{(a,b)}_x$, 
\be
q^{(a,b)}_x = \rep{a}_x\rep{b}_x\,\in \Z_2
.\ee
The probability distribution of the overlap fields, $\mu$,
can be implicitly defined 
through the overlap expectations, that involve both the thermal
average and the average $E$ over disorder. For any smooth function $F$
define 
\be
\dlangle F\left( q^{(12)}_x, q^{(23)}_y,\ldots\right) \drangle
= E\left[ \langle F\left( \rep{1}_x\rep{2}_x,\, \rep{2}_y\rep{3}_y,
\ldots\right)
\rangle\right],
\label{overlap-av}
\ee
where expectation with respect to the $\mu$
distribution is denoted by $\dlangle\cdot\drangle$.  
All physical observables can be expressed in terms of overlap
observables, so that the full physical meaning of these models is
contained in the overlap probability measure.

Let us introduce the volume average of the overlap field,
the total overlap, often simply referred to as ``overlap'', 
\be
q^{(a,b)} = \frac{1}{|\Lambda|} \sum_x q^{(a,b)}_x
.\ee
For two replicas, the distribution of the overlap $q^{(1,2)}$ will
be denoted by $P^{(2)}$. 
An other interesting observable is the two points, connected
correlation function of the overlap field: 
\be
C(r)=\frac{1}{|\Lambda|} \sum_x\dlangle
q_x^{(1,2)}q_{x+r}^{(1,2)}\drangle_c
,\label{corr}\ee 

\section{A renormalization group for spin glass models}
\label{sec:RG}

The total overlap is the order parameter of the
model. In the high temperature phase, including the critical point, 
it should be zero in the
infinite volume limit, while in the low temperature phase 
it is expected to fluctuate~\cite{MPV,Marinari:1999a,droplet}. 
Therefore, the critical point can be characterized by a divergence of the correlation
length associated to \Ref{corr}. For this reason, it is natural to
define the RG transformation on the overlap field.
In the following we consider only two replicas, and omit the replica
indexes on the overlap variables. \par 
Let $B^n_x \subset \Lambda$ be the cube of size $n$ located in $x$ and
$\Lambda_n\subset\Z^d$ a cubic lattice of side $L/n$. Because of
$\Z_2$ symmetry, the coarse grained
field $q'_{x'}$ can be defined~\cite{Wang:1988} on $\Lambda_n$
as the sign of the majority of overlaps inside a block centered on $x=nx'$.
The renormalization group transformation for a rescaling factor $n$
then acts on the overlap distribution $\mu$ as follows:
\be
\mu'_{n} (q') = \sum_{\{q\}} \mu (q)\prod_{x' \in \Lambda_n} \delta
\left(q'_{x'}\,, \sign \left(\sum_{y \in B^n_{nx'}} q_{y}
    \right)\right),
\label{rg-trans}
\ee
where the sum on the r.h.s. runs over all the $2^{|\Lambda|}$ 
configurations of overlaps $q_x$ and 
$\delta(a,\,b)$ is 1 for $a=b$ and zero otherwise. 
For $m=n^k$, one defines the distribution $\mu_n^{(k)}(q^{(k)})$ as the result
of $k$ iterations of \Ref{rg-trans}, and the semi-group property
holds by construction. \par
Extending these definitions, in the appropriate way, to the
thermodynamic limit, the transformation may be iterated indefinitely
and eventually a non-trivial asymptotic distribution $\mu^\ast$ may be
reached. 
However, to carry out useful RG calculations an approximation scheme must be
chosen. In this work we have used the Monte Carlo approach, for which
the main approximation is the use of a finite lattice.
The basic effect is that of neglecting the influence of distant regions of
the lattice on each other in the calculation of $\mu'_n(q')$
from $\mu(q)$~\cite{BellWilson:finite}. Indeed, a fundamental property of a RG transformation is that
the parameters representing the renormalized distribution must depend
analytically on the unrenormalized parameters. 
One assumes that
the dependence of $\mu'_n$ on $\mu$ is analytic, because local
observables of $q'$ in one region of the lattice are not
appreciably affected by distant regions of the system, in spite of the
eventually infinite correlation length. 
For the very same reason, the computation of local observables of the
field $q'$ should be weakly affected by finite lattice effects, on
the contrary of what happens to observables of the unrenormalized
field $q$.  \par
On a finite lattice the transformation~\Ref{rg-trans} may be limited to a very small number
of iterations. Another typical
problem is the truncation error due to the projection of renormalized
distributions onto a space with a restricted number of couplings. To
minimize the effects of these approximations, 
a different RG scheme~\cite{BP89,NOI} has been adopted. 
We define the renormalized fields $q'_x$ on
$\Lambda_n$ as given just by
the transformation~\Ref{rg-trans} with $n=L/\ell$ (with small typical
values, \eg~$\ell = 2,3$); the renormalized
distribution will be denoted by $\mu'(q')$.
Notice that, with only one iteration, $\Lambda_n$ is a lattice with
volume $\ell^3$.
Due to the smallness of $\Lambda_n$ only a few
couplings are allowed and there is no truncation error. The procedure
is repeated for different values $L,L'$ and the ratio $L/L'$ is
interpreted as a scaling factor~\cite{BP89} (but see also~\cite{Night}). 
Moreover, the sequence will give better results for increasing $n$,
because the group parameter will be large and, at the same time, the
size of the initial system will be affected by smaller finite-size
effects.
Given the large size of blocks,
we also expect the results to be not dramatically dependent on the
choice of blocking transformation. 

The RG transformation is naturally defined on the probability
distribution of the overlap field. For the purpose of calculations this
distribution is well characterized by the expectations of a set of
(translation-invariant) observables. It is the approach that has
been followed in this work. However, the question may be raised of
how to express such a distribution in terms of physically meaningful
parameters. It is clear that the traditional exponential form,
$\exp(-H)$, has no physical relevance, due to the involved definition of the
overlap field distribution in the original model. To state the
question in a different way: which additional interactions between the
microscopic variables are generated as a result of the RG transformation~?
The answer to this question is important in order to get high
precision estimates of the critical indexes, through the observation
of the effects of irrelevant perturbations to the fixed point.  
We propose the following parameterization, that proved to be effective
in the case of dimension $d=4$~\cite{NOI}. The Hamiltonian, given
by \Ref{hamiltonian}, is kept fixed, while the disorder is distributed
according to a general gauge action:
\be
\rho_K(J) = \exp\left( {\sum}_i K_i \, W_i(J) \right)
\label{gauge-disorder}
,\ee
where the $K_i\in\R$ are parameters and the $W_i$'s are Wilson's
loops, \ie,  products of $J$'s along closed paths. Let us stress that
the effects of this parameterization are that
spins get non trivial extra correlations through the
``dressing'' of interactions.  
The resulting set of spin models will be discussed in more detail in
next section
of this paper.


\section{Spin models with quenched gauge-field interactions}
\label{sec:universality}
 
Consider the E-A spin Hamiltonian \Ref{hamiltonian}, and assume the
quenched interactions are distributed according to a general,
gauge-invariant distribution function such as \Ref{gauge-disorder}.
A fairly large set of models may be defined in this way.
For the sake of simplicity, let us restrict our attention to the
disorder distributions of the following form,
\be
\rho_K (J) = C_K\,\exp\left( 
K_1\sum_{\langle x,\,y\rangle} J_{xy}^2
+ K_2\sum_{\langle x,\,y\rangle} J_{xy}^4
+ K_3 \sum_\alpha \plaq_\alpha  
\right)
,\label{J_K1K2K3}\ee
where $C_K\in\R$ is a normalization constant, and the symbol $\plaq$ 
denotes the plaquette terms of the kind $J_{x,y} J_{y,z} J_{z,w} J_{w,x}$. 
Expectation with respect to this distribution will be denoted by $E_K$.
We notice that the Gaussian Edwards-Anderson model corresponds to $K_1
= -1/2,\ K_2= K_3 =0$, while the $\Z_2$, E-A model is obtained in the
limit $K_1, K_2 \rightarrow \infty,\ \mathrm{s.t.}\ K_1/K_2 = -2,\
K_3 = 0$.
But a more interesting model is obtained by adding a plaquette term
(\ie, $K_3\not=0$): 
\bea
\rho_K (J) = C_K\, e^{K_3 \sum_\alpha \plaq_\alpha},\;\;
J\in \Z_2^{d|\Lambda|}
,\label{pure_gauge}
\eea
Considering the disorder variables only, this is the well known pure
gauge $\Z_2$ model~\cite{Balian:1975ir}. The partition function, the free energy and the internal energy density are defined by
\bea
& Z_\Lambda(K_3) = \sum_{\{J_{xy}=\pm 1\}} \exp\left( K_3 \sum_\alpha
\plaq_\alpha \right) = e^{- K_3 F_\Lambda(K_3)} 
&\\
& u_\Lambda (K_3) = - E_{K_3} \left( \sum_\alpha
\plaq_\alpha / N_\plaq \right) 
.&\eea
The energy density $u_\Lambda$ is equal to minus the average plaquette, thus it
gives a measure of the short -- range frustration of the spin system. 

Let us denote by $w_\gamma$ the product of the $J$'s along a closed path $\gamma$,
\be
w_\gamma = \prod_{\gamma} J_{xy} = \prod_{\alpha\in S} \plaq_\alpha
,\ee
where $S$ is a surface bounded by $\gamma$.
Two canonical asymptotic regimes may be distinguished:
a {\it weak decrease} regime, characterized by $E ( w_\gamma ) \approx
e^{-L}$;
a {\it strong decrease} regime, where $E ( w_\gamma ) \approx e^{-A}$.
$L$ and $A$ denote the perimeter of $\gamma$ and the area of the surface $S$.
In various models a transition from the strong to the weak decrease is
observed as the average plaquette increases.  
The model \Ref{pure_gauge}, in 3 dimensions, exhibits a second order
phase transition at $K_3=K_3^c \simeq 0.7613$. The corresponding order
parameter is the square Polyakov loop, $p^2=(p_1^2+p_2^2+p_3^2)/3$,
where $p_i$ is the average Wilson loop along a path that closes itself
exploiting the periodicity of the b.c.'s in direction $i$.  
The quantity $E(w_\gamma)$ is relevant for the spin system associated
to the gauge field~\cite{Toulouse:1977},
as it is related to the average frustration on the range $\gamma$. 
%

We are now in the position to sketch the behavior of the spin system
corresponding to the pure-gauge disorder distribution~\Ref{pure_gauge}
\cite{NOI}:  
\begin{description}
\item[$K_3\rightarrow\infty$] The disorder is in a ground state of the
  gauge Hamiltonian, characterized by $E_{K_3}(\plaq)=1$, $p_i=\pm
  1$. Up to a gauge transformation, the spin model is an Ising model,
  with periodic b.c.\ in the directions $i$ for which $p_i=1$ and
  anti-periodic in the others. 
\item[$K^c<K_3<+\infty$] The disorder is in the weak decrease phase: 
  frustration effects are very small
  locally ($E_{K_3}(\plaq)\approx 1$) and increase weakly at long
  range. The spin system is expected to be ferromagnetic.
\item[$-K^c<K_3<K^c$] The disorder is in the strong decrease phase. 
  The average plaquette is approximated by   
  $E_{K_3}(\plaq)\approx K_3$, except in the region very close to the 
  transition. However, frustration moves rapidly toward $1/2$ for
  increasing range. Spin glass behavior is expected.  
\item[$K_3\rightarrow -\infty$] The gauge field is in a ``fully
  frustrated'' configuration, \ie,  all plaquettes are
  frustrated. 
\end{description}
The behavior in the region $-\infty<K_3<K^c$ is still an open question.

These different qualitative regimes should correspond to
different universality classes of the spin system.
Exploiting the RG approach introduced in the previous
section, we will be able to characterize the universality classes in terms 
of different fixed points. 


\section{Analysis of Monte Carlo computations}
\label{sec:num}

The Renormalization Group analysis described in section~\ref{sec:RG}
was performed numerically by dynamic Monte Carlo computations. 
We employed Hamiltonian \Ref{hamiltonian} in dimension $d=3$, with
periodic boundary conditions and $\Z_2$
quenched disorder distributed according \Ref{pure_gauge}.
Besides the traditional Edwards-Anderson model, corresponding to the
 case $K_3=0$, we simulated two other models, with $K_3=0.3$
and 0.8, in order to verify the conjecture on universality classes
made in section~\ref{sec:universality}. Moreover,
the MC-RG analysis was performed on the Ising model, corresponding to
$K_3=\infty$, as a test of the method, and for a comparison to the
other models. 

The choice of $K_3$ values deserves some comments. $K_3=0.3$ is far
from the critical value $K_3^c \simeq 0.7613$ to avoid cross-over
effects, and is quite different from the E-A model: frustration is
$f_{0.3}\simeq 0.35$, compared to $f_0=0.5$. In the deconfined phase, $K_3=0.8$  
gives a very low frustration $f_{0.8} \simeq 0.015$. Lower values of $K_3$
provide 
slightly higher frustration, but we considered unwise a value too
close to $K_3^c$, even though the critical coupling is lower in a
finite lattice. Nevertheless, the model is quite different from the
Ising model, as is shown, for instance, by the value of the square
Polyakov loop, $p^2 \simeq 0.65$. 
Among the eight possible pure phases, we chose the one
characterized by positive Polyakov loops, \ie, $p_i\geq 0$. 

For each model, a quite large number of samples (with different configurations of the
disorder $J$) were simulated, as reported in table~\ref{MCtable}. 
The required disorder configurations were generated by an independent
MC process on the gauge fields, according to distribution~\Ref{pure_gauge}.
This step is quite simple with nowadays computers. The average
correlation between $J$ configurations is never greater than
$10^{-4}$. 

The MC runs on the spin variables are more complicated. We simulated
lattices of linear size $L= 8,\,12,\,16$ for all models, together with
additional $L=24$ for models $K_3=0.8$ and Ising and
 $L=10,\,14$ for $K_3=0$ whose significance will be clear later.  
The code makes use of
Parallel Tempering (PT) algorithm and multi-spin encoding.
For every Metropolis update, a PT step was performed. 
Two replicas were simulated through
independent MC chains in order to compute overlap observables. 
The RG transformation was carried out following the procedure 
explained in section~\ref{sec:RG} \cite{BP89}.
Therefore, the RG parameter $n$ takes the values 4, 6 and 8
(and 12) as well as 5, 7 for $L=10,\,14$. 
To characterize the distribution $\mu'$, we measured the
following observables of the random field $q'$:
\begin{eqnarray}
A_1  = \frac{1}{3|\Lambda'|} \sum_{x,l:|l|=1} \dlangle q'_x q'_{x+l} 
\drangle \hspace{+6ex} &
A_4 & \hspace{-2ex} = \dlangle S^2 \drangle  \\
A_2  = \frac{1}{3|\Lambda'|} \sum_{x,l:|l|=\sqrt{2}} \dlangle q'_x q'_{x+l} 
\drangle  \hspace{+6ex} &  
A_5 & \hspace{-2ex} = \dlangle S^4 \drangle \\
A_3  = \frac{1}{3|\Lambda'|} \sum'_{x,y,z,w} \dlangle  q'_x q'_y q'_z q'_w
\drangle \hspace{+6ex} &  
A_6 & \hspace{-2ex} = \dlangle ( 0.4 + S^2 )^{-1} \drangle   
\end{eqnarray}
where $S= |\Lambda'|^{-1} \sum_x q'_x$, and the sum in $A_3$ runs over
the 4-uples of sites located on plaquettes. Notice that observable
$A_6$ is sensible to the small values of $S$, as opposed to
$A_4$ and $A_5$, and 0.4 is an arbitrary cut-off. 
For the sign function involved in the definition of $q'$, a tie-breaker
was used in the undetermined zero cases. Models with $L=10,\,14$ are  
unaffected by this procedure and, in principle, this difference in the 
transformation rule could provide different results.   
Moreover, we have explicitly measured the $\beta$
derivative of each observable, computing the connected correlation
function with the Hamiltonian.  
In the definition of the blocks, the choice of the origin is
arbitrary. In order to improve the quality of statistics, 
we choose to average the measures over eight different origins. 
The lower simulated temperature, $T_\mathrm{min}$ in
Table~\ref{MCtable}, was chosen around $0.9T_\mathrm{c}$, according to previous
estimates of $T_\mathrm{c}$, for the $K_3=0$ model. 
The efficiency of the PT algorithm was checked through the acceptance
rate of temperature swaps. We ensured it was roughly 1/2 in the
most difficult runs, and never below 40\%. 

In the rest of this section we discuss the checks on ergodicity of MC
runs and error estimates. Generally speaking, as higher
values of $K_3$ correspond to lower frustration, 
systems should get ordered more easily so
that we expect critical temperatures to increase with $K_3$,
carrying on a global improvement of MC-PT performances. 
On the opposite, negative values of $K_3$ presumably give rise to very
low critical temperatures thus making simulations very difficult, at
least in dimension $d=3$.
A well known problem in performing Monte Carlo computations on glassy
systems is the difficulty in reaching ergodic regime, due to strong
metastability effects and to the large number of nearly degenerate
ground states.
A very powerful --- albeit indirect --- criterion to ensure ergodicity
makes use of the observation of (Monte Carlo) time spent by the system
at each temperature of the PT set. This must be approximately the same
for each temperature. We measured the ratio $r$ of the time spent in the
most visited T over the less visited one. The histograms of the
results are presented in Table~\ref{MCtable}.
\begin{table}
\begin{center}
\begin{tabular}{lrcrrlllllll}
\hline \noalign{\smallskip}
$K_3$ & $L$ & $N_\mathrm{_{MC}}/10^5$ & $N_\mathrm{S}$ & $N_{\beta}$
& $T_\mathrm{min}$ & $T_\mathrm{max}$ & $r\geq\sqrt{2}$ & $r\geq 2$ & $r\geq 4$\\
\noalign{\smallskip}\hline\noalign{\smallskip}
0   &  8 & 1.0 & 36096 & 16 & 1.00 & 2.30 & 0    & 0     & 0    \\
    & 10 & 1.2 & 10240 & 19 & 1.00 & 2.35 & 0.02 & 0     & 0    \\
    & 12 & 1.6 & 15936 & 27 & 1.00 & 2.30 & 0.18 & 0.005 & 0    \\ 
    & 14 & 2.0 & 5632  & 30 & 1.01 & 2.32 & 0.38 & 0.04  & 0    \\
    & 16 & 2.4 & 3328  & 33 & 1.02 & 2.30 & 0.50 & 0.09  & 0.001 \\ 
\hline 
0.3 &  8 & 1.0 & 20160 & 26 & 1.20 & 3.00 & 0    & 0     & 0     \\ 
    & 12 & 1.0 & 9216  & 36 & 1.10 & 2.85 & 0.55 & 0.12  & 0.005 \\ 
    & 16 & 2.0 & 2880  & 45 & 1.20 & 2.96 & 0.84 & 0.40  & 0.08  \\
\hline 
0.8 & 8  & 0.5 & 4000  & 11 & 4.00 & 5.00 & 0    & 0     & 0  \\ 
    & 12 & 0.2 & 4000  & 20 & 4.05 & 5.00 & 0    & 0     & 0  \\ 
    & 16 & 0.5 & 512   & 28 & 3.92 & 5.00 & 0.37 & 0.02  & 0  \\ 
    & 24 & 0.5 & 512   & 9  & 4.44 & 4.52 & 0    & 0     & 0  \\
\hline 
\end{tabular}
\end{center}
\caption{\textit{Parameters of MC-PT runs. The same number of MC-PT
steps, $N_\mathrm{MC-PT}$, was used for thermalization and measurements. 
We report also the number of different disorder samples $N_\mathrm{S}$, the
total number of temperatures 
    $N_{\beta}$ (together with minimum and maximum values) for the PT
    algorithm, and the histograms on visiting frequencies.}}  
\label{MCtable}
\end{table}

\begin{figure}
\resizebox{0.5\textwidth}{!}{\includegraphics{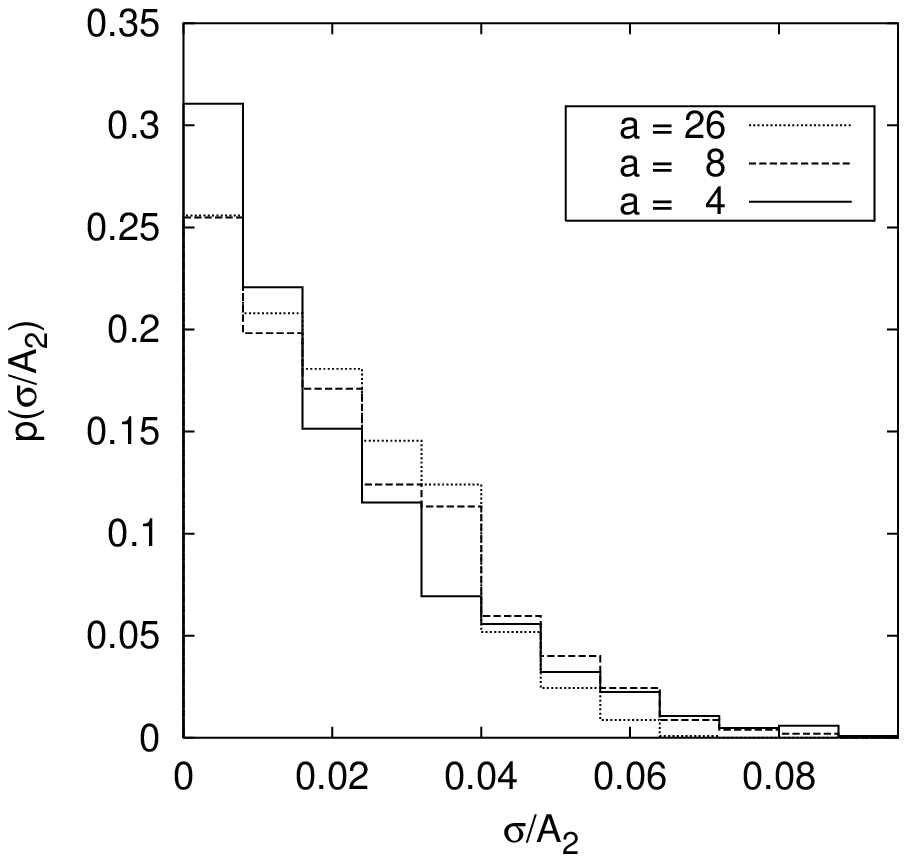}}
\resizebox{0.5\textwidth}{!}{\includegraphics{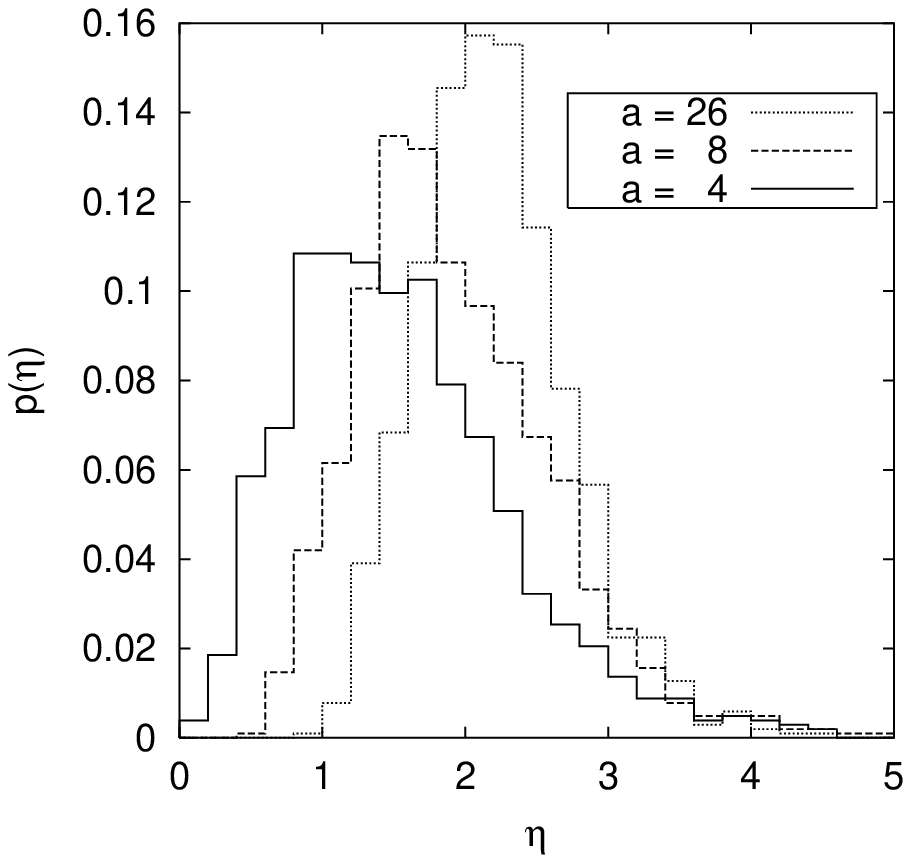}}
\caption{\textit{Relative error (on the left) and ergodicity
    distribution (right) of observable $A_2$ on various time scales
    $l/a$ in the E-A model $K_3=0,\ L=16,\ T=1.02$.
    The total number of
    samples examined is 1024. Relative error diagram is quite stable
    and more than 99\% of sample have relative errors lower than 0.08,
on all time scales. On the right, the distribution of $\eta$ shows
    that on short time scales ($a = 26$), 63\% of samples have $\eta
    > 2$; this number drops to 41\% for $a=8$, and to 25\% for $a=4$, 
    indicating run
    length is essentially correct. Single sample errors and ergodicity
both improve at higher temperatures.}} 
\label{fig:er}
\end{figure}

Ergodicity may be verified explicitly by analyzing the
time sequence of the measures. By the same token we can estimate
the statistical error of the measures taken on a single sample. 
The algorithm employed measures
observables at each MC-PT step. These measures were averaged in groups
of 1024, providing a set of a few hundred points per sample and per
temperature. One problem of PT algorithm is that of properly defining
an autocorrelation time.
However, these points may be considered reasonably independent from
each other. Thus, we can estimate the statistical error on a single
sample. Then the procedure is repeated: the measures are averaged
over groups of points, obtaining a different estimate of the error.
Clearly, as the size of the groups increases, the resulting points will
be less correlated producing a better (usually larger) estimate of the
error. In Figure~\ref{fig:er} (left), we plotted the histogram of the
single-sample relative errors on observable $A_2$, 
in our worst-case simulation ($K_3=0,\ L=16,\ T=1.02$). The quantity
$l$ denotes the number of initial points and $a$ is the number of groups 
so that the size of groups is roughly $l/a$. The results
seem to be very stable in $a$. 

We can check ergodicity in a similar way. On each group
$i=1,\,\ldots,\, a$ we
compute, along with the average observable $x^{(a)}_i$, 
the error on the average, \ie, its
standard deviation (denoted by $\tau^{(a)}_i)$. 
If ergodicity is reached on a time scale $l/a$ then the $x^{(a)}_i$'s
will be nearly aligned and fluctuations around the mean will
be of order $\sim\bar{\tau}=a^{-1}\sum_i \tau_i^{(a)}$. 
On the other hand, if $l/a$ is a
too short time scale, the points will spread with
very large fluctuations around the mean. Let us denote by $\sigma_a$
the standard deviation of group averages $x^{(a)}_i$.
To characterize ergodicity we thus measure the ratio
$\eta=\sigma_a/\bar{\tau}$, expecting $\eta \stackrel{<}{\sim} 1$ when
ergodicity is reached, and 
$\eta \gg 1$ when is not. The histograms of the results for
different samples are shown in Figure~\ref{fig:er} (right).
Let us underline that the results should be extrapolated to $a=1$:  
a ``good'' MC run should not be much longer than necessary.

From general arguments, the total statistical error on 
observable $A$, $\Delta A$, can be estimated as follows:
\begin{equation}
(\Delta A)^2 = \frac{\epsilon_S^2 + \sigma^2}{N_{S}}
\label{eq:err}
\end{equation}
where $\epsilon_S^2$ is the sample-to-sample variance, $N_\mathrm{S}$ the total
number of simulated samples and $\sigma$ is the single-sample
error. As typical sample-to-sample fluctuations of observables 
are of order 100\%, 
the single-sample errors can be neglected.

\section{Numerical results}

Performing the Renormalization Group analysis explained in
section~\ref{sec:RG}, we are able to characterize universality
classes depending on frustration parameter, pointing out the influence
of deconfinement phase transition of gauge fields on the behavior of
the associated spin model, namely in the transition from glassy to
ferromagnetic behavior.
In particular the results confirm the
conjecture~\cite{Toulouse:1977} that the asymptotic regime of long range
frustration is responsible for the glassy behaviour.
  Universality classes were
determined by characterizing the fixed point, therefore on
fundamental grounds, without relying on phenomenological comparison
of critical indexes. 
Moreover, the RG analysis provided good
estimates of the critical temperature, $T_c$, and thermal index $\nu$,
of the four models considered: $K_3=0,\, 0.3,\, 0.8$ and Ising. 
As we will show in this section, the fundamental determination of
universality classes allows to consider $\nu_0$ and $\nu_{0.3}$ as
\emph{independent estimates} of the thermal index of the glassy fixed point.

In order to damp irrelevant perturbations, and to enhance the effect
of the relevant coupling~\cite{BP89}, we chose to employ the largest possible
values of $n = L/L'= 4,\, 6,\, 8,\, (12)$, thus setting the
renormalized lattice size equal to $L'=2$.  
As a drawback, observables $A_1,\,\ldots,\,A_6$ turned out 
to be strongly correlated, making it impossible to perform a full
linearization of the RG transformation on the fixed point and to
estimate irrelevant exponents. 

\begin{figure}
\resizebox{\textwidth}{!}{\includegraphics{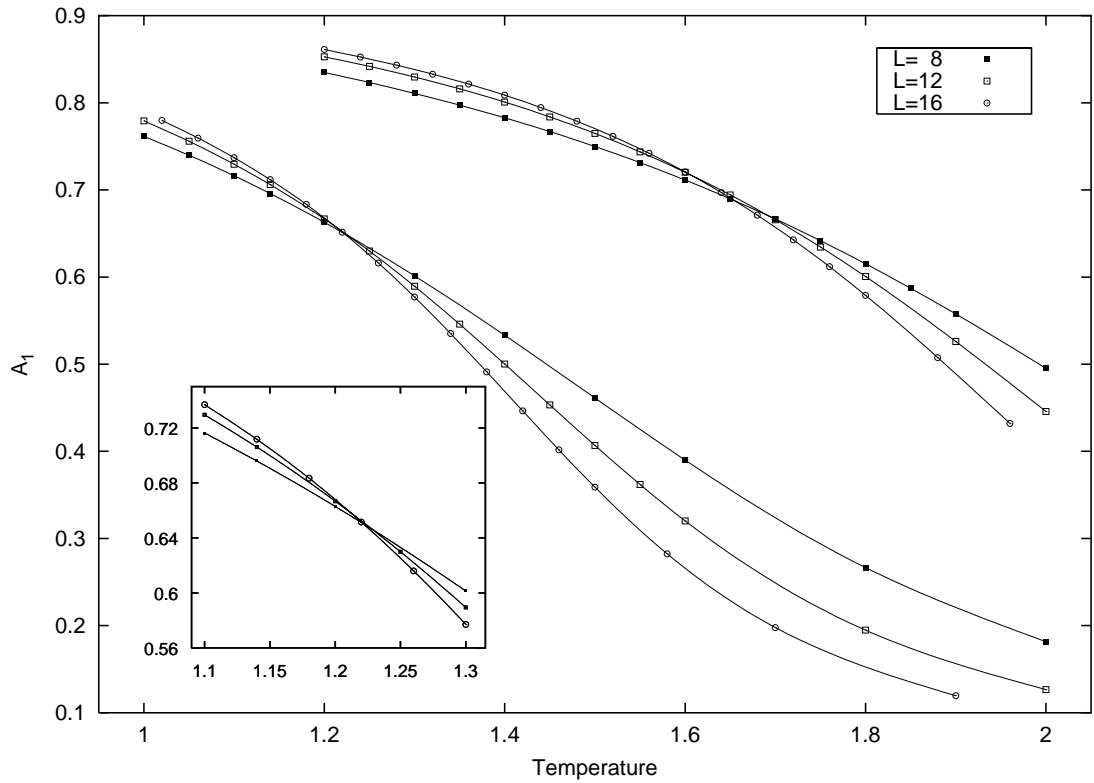}}
\caption{\textit{Observable $A_1$ versus temperature for E-A model
$K_3 = 0$ (on bottom, with a magnified detail) 
and for model $K_3 = 0.3$ (top). Points size is larger than error-bars,
except in the detail.}}
\label{fig:crossa0a300}
\end{figure}

\begin{figure}
\resizebox{0.5\textwidth}{!}{\includegraphics{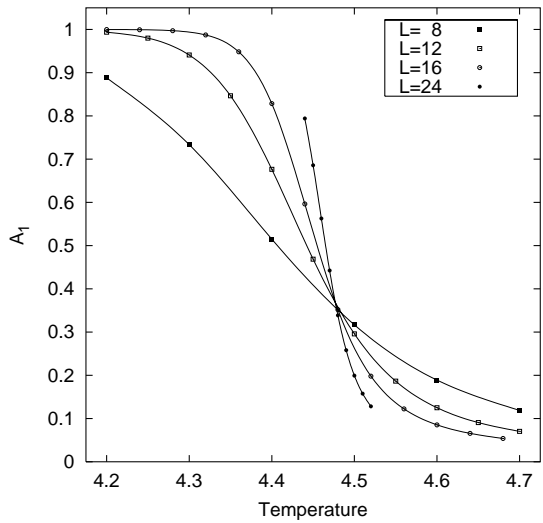}}
\caption{\textit{Observable $A_1$ versus temperature for $K_3 = 0.8$.}}
\label{fig:crossa800}
\end{figure}

Observables defined on different lattices characterize the
trajectory of RG flow and, at the critical temperature, 
they move toward a fixed point where they become independent of
the initial lattice. As a 
function of temperature, observables will cross in a point which can
be estimated as the critical temperature $T_c$. 
The linearization around the crossing point of two quantities $A(L_1)$ and
$A(L_2)$ determines the thermal exponent through:  
\begin{equation} 
\label{eq:nu}
 \nu^{-1}(A) = \frac{\ln \left[ \frac{dA(L_1)}{d\beta}|_{\beta_c} \; / \; {\frac{dA(L_2)}{d\beta}|_{\beta_c}} \right] }{\ln (L_1 /L_2)}
\end{equation}
where $\beta^{-1} = T$ and derivatives are computed at the
critical point. In Figures~\ref{fig:crossa0a300} and \ref{fig:crossa800} is
plotted observable $A_1$ as a function of temperature, the points size
being larger than the error bars, except in the detail of
Figure~\ref{fig:crossa0a300}.
In accordance with \cite{Ballesteros:2000}, Figure~\ref{fig:crossa0a300}
seems to rule out a Kosterlitz-Thouless transition type, for in that
case the curves should merge under $T_c$. Moreover, the resulting estimates of
$\nu$ obtained with formula \Ref{eq:nu} are the best proof that curves
cross with different slopes.

Data, for each $A_i$, were fitted to allow for extrapolation to the critical point. 
We notice incidentally that the measures of
derivatives with respect to $\beta$ greatly improved the precision of the results.
The errors on $T_c$ and $\nu$ were estimated with the following
procedure: measures taken on different samples were grouped in 32
blocks and averaged over 31 blocks 
to compute the results. The procedure was repeated with
different blocks, obtaining 32 different estimates of $T_c$ and $\nu$.
Then, errors were estimated with jackknife formula.

The results shown in Table~\ref{tab:ind} were averaged over the six
observables. 
Notice that, for glassy models, only two of the three estimates
provided are independent, as they all come from the same three lattices.
The error on $T_c$ and, as a consequence, on $\nu$ is larger for 
the pair $(16,\,12)$ because of the very close slopes of curves.
A transient effect due to the flow of irrelevant couplings seems to
be present, but cannot be measured within current precision and
lattice sizes. Results for $K_3=0.8$ have much higher precision,
revealing that sample-to-sample fluctuations are very small.
The analysis of the Ising model was performed to serve as a benchmark of the
method. Recent high precision estimates, in the infinite volume limit, are
as follows~\cite{Blote:1999,Pellis}: $T_c= 4.511524(2),\; \nu=0.63012(16)$. 
The results for $K_3=0.8$ and Ising confirm the effectiveness of the
method, and suggest that the evaluation of irrelevant exponents is   
necessary in order for to obtain reliable results at higher precision.
Notice the results from lattices $(14,\,10)$:
in this case a different transformation $\Theta$ was used 
(the tie-breaker is not necessary).
Our results for the Edwards-Anderson model are in agreement with
previous estimates~\cite{Palassini:1999,Ballesteros:2000}, obtained
with finite-size scaling techniques on much larger lattices and
exploiting dedicated machines or supercomputers.

Finally, we present in Table~\ref{tab:fixp} the
values of observables
$A_1,\ldots,A_6$ at the estimated critical point. These data roughly
characterize the fixed point on the lattice $L'=2$.
Remnant effects of irrelevant couplings affect the
determination of $T_c$, and therefore our extrapolation.
Nevertheless, data points clearly point out the existence of 
different universality classes below and above the deconfinement
transition. Within current precision, we can very well assume that
model $K_3=0,\, 0.3$ are in the same \emph{spin glass} universality
class, while $K_3=0.8$ appears to be in the same class as Ising model.
\begin{table}
\begin{center}
\begin{tabular}{llll}
\hline\noalign{\smallskip}
$K_3$ & $(L_2,\,L_1)$ & $T_\mathrm{c}$ & $\nu$ \\  
\noalign{\smallskip}\hline\noalign{\smallskip}
0 & (12,\,8) & 1.225 (10) & 1.89 (9) \\
  & (16,\,8) & 1.221 (12) & 1.96 (12) \\
  & (16,\,12) & 1.215 (25) & 2.11 (34) \\
\hline 
0 & (14,\,10) & 1.186 (36) & 2.12 (36) \\
\hline 
0.3 & (12,\,8)  & 1.732 (25) & 1.87 (13) \\
    & (16,\,8)  & 1.681 (20) & 1.87 (11) \\
    & (16,\,12) & 1.62 (6)   & 1.9 (5) \\
\hline 
0.8 & (12,\,8)  & 4.4833 (6) & 0.6637 (8) \\
    & (16,\,8)  & 4.4793 (6) & 0.6563 (10) \\
    & (16,\,12) & 4.4761 (13)& 0.652  (3) \\
    & (24,\,8)  & 4.4779 (2) & 0.6512 (3) \\
    & (24,\,12) & 4.4767 (3) & 0.6439 (5) \\
    & (24,\,16) & 4.4769 (7) & 0.6395 (12)\\
\hline 
$\infty$ & (12,\,8)  & 4.5213 (2) & 0.6588 (9) \\ 
         & (16,\,8)  & 4.5178 (1)  & 0.6537 (7) \\
         & (16,\,12) & 4.5149 (2) & 0.648 (2) \\
         & (24,\,8)  & 4.5149 (1)  & 0.6495 (5) \\ 
         & (24,\,12) & 4.5134 (1)  & 0.6441 (8) \\
         & (24,\,16) & 4.5128 (1)  & 0.6417 (14) \\
\noalign{\smallskip}\hline
\end{tabular}
\end{center}
\caption{\textit{Estimates of the critical temperature ($T_\mathrm{c}$) and the
    critical index $\nu$. $K_3=\infty$ denotes the Ising model (with
    periodic b.\ c.).} \label{tab:ind}} 
\end{table}

\begin{table}
\begin{center}
\small
\begin{tabular}{llllllll}
\hline\noalign{\smallskip}
$K_3$ & $(L_1,\,L_2)$      & $A_1$    & $A_2$    & $A_3$    & $A_4$    & $A_5$    & $A_6$ \\  
\noalign{\smallskip}\hline\noalign{\smallskip}
0 & $(12,\,8)$   & 0.647 (6) & 0.580 (8) & 0.520 (7) & 0.653 (7) & 0.570 (8) & 1.143 (10) \\
  & $(16,\,8)$   & 0.650 (7) & 0.583 (9) & 0.523 (8) & 0.656 (8) & 0.573 (9) & 1.139 (11)  \\
  & $(16,\,12)$ & 0.656 (18)& 0.589 (23)& 0.528 (20)& 0.660 (19)& 0.578 (21)& 1.133 (27)  \\
\hline 
0.3 & $(12,\,8)$   & 0.645 (14) & 0.574 (17) & 0.512 (16) & 0.649 (14) & 0.562 (15) & 1.145 (20) \\
    & $(16,\,8)$   & 0.670 (10) & 0.603 (12) & 0.543 (12) & 0.673 (10) & 0.592 (11) & 1.114 (14) \\
    & $(16,\,12)$ & 0.71 (3)   & 0.65 (4)   & 0.59 (4)   & 0.71 (3)   & 0.64 (4)   & 1.07 (4) \\
\hline 
0.8 & $(12,\,8)$   & 0.3429 (14) & 0.2920 (15) & 0.1840 (11) & 0.3969 (13) & 0.2805 (13) & 1.5103 (21) \\
    & $(16,\,8)$   & 0.3505 (12) & 0.2990 (13) & 0.1924 (11) & 0.4032 (11) & 0.2877 (11) & 1.5012 (19) \\
    & $(16,\,12)$ & 0.369 (5)   & 0.316 (5)   & 0.213 (4)   & 0.419 (4)   & 0.305 (5)   & 1.479 (7) \\
    & $(24,\,16)$ & 0.367 (6)   & 0.310 (6)   & 0.215 (5)   & 0.415 (5)   & 0.303 (5)   & 1.489 (9) \\
\hline 
$\infty$ & $(12,\,8)$   & 0.3390 (4) & 0.2881 (4) & 0.1830 (3) & 0.3936 (4) & 0.2780 (4) & 1.5166 (6) \\
         & $(16,\,8)$   & 0.3460 (3) & 0.2945 (4) & 0.1899 (3) & 0.3993 (3) & 0.2842 (3) & 1.5078 (5) \\
         & $(16,\,12)$ & 0.3629 (6) & 0.3099 (8) & 0.2068 (6) & 0.4133 (6) & 0.2994 (6) & 1.4869 (11) \\
         & $(24,\,16)$ & 0.3757 (8) & 0.3207 (9) & 0.2202 (7) & 0.4234 (8) & 0.3108 (8) & 1.4726 (12) \\

\noalign{\smallskip}\hline
\end{tabular}
\normalsize
\end{center}
\caption{\textit{Characterization of the fixed points: observables
    $A_1,\ldots,A_6$ computed on the lattice $L_2$ at the estimated
    critical temperature $T_c(L_1,\,L_2)$.}}\label{tab:fixp} 
\end{table}

\section{Conclusions}

A recently introduced~\cite{NOI} (see however~\cite{Wang:1988}) 
Renormalization Group transformation suitable for disordered spin
models is explained in detail and applied numerically in the three
dimensional case.  
Universality classes depending on a frustration parameter have been
characterized, pointing out the influence
of deconfinement phase transition of gauge fields on the behaviour of
the associated spin model, namely in the transition from glassy to
ferromagnetic behavior.
The study of universality classes relies on the direct detection
of fixed points, therefore on
fundamental grounds, without relying on phenomenological comparison
of critical indexes.   
Moreover, the RG analysis provided good
estimates of the critical temperature, $T_c$, and thermal index $\nu$,
of the four models considered: $K_3=0,\, 0.3,\, 0.8$ and Ising. 
As we have shown, the fundamental determination of
universality classes allows to consider $\nu_0$ and $\nu_{0.3}$ as
\emph{independent estimates} of the thermal index of the glassy fixed point.
These results are in agreement with previous numerical estimates on
the Edwards-Anderson model, obtained exploiting Finite-Size Scaling 
Ansatz~\cite{Palassini:1999,Ballesteros:2000,MariCampbell}.
For the small lattices and the reasonable computer time employed, this
RG technique is to be considered very effective in order to compute
critical properties of disordered systems.

\section*{Acknowledgments}
The authors wish to warmly thank Giorgio Parisi for interesting
discussions.

\newcommand{\PRL}{Phys.\ Rev.\ Lett.\ }
\newcommand{\PR}{Phys.\ Rev.\ }
\newcommand{\CP}{Comm.\ Phys.\ }
\newcommand{\CMP}{Comm.\ Math.\ Phys.\ }
\newcommand{\JSP}{J.\ Stat.\ Phys.\ }
\newcommand{\PyL}{Phys.\ Lett.\ }
\newcommand{\JMM}{J.\ of Magnetism and Magnetic Materials\ }
\newcommand{\IJMP}{Int.\ J.\ Mod.\ Phys.\ }
\def\JdP{J.\ Physique\ }
\newcommand{\JP}{J.\ Phys.\ }
\def\EPJ{Eur.\ Phys.\ J.\ }
\newcommand{\EL}{Europhys.\ Lett.\ }

\end{document}